\documentclass[twocolumn]{aastex631}

\usepackage[utf8]{inputenc}
\usepackage[T1]{fontenc}
\usepackage{mathptmx}
\usepackage{etoolbox}
\usepackage{amsmath}
\usepackage{amssymb}

\def \sh {\text{sh}}
\def \f {\text{f}}
\def \p {\text{p}}
\def \e {\text{e}}
\def \s {\text{s}}
\def \d {\text{d}}
\def \u {\text{u}}

\def \i {\text{i}}

\def \y {\text{y}}

\def \L {\text{L}}
\def \E {\text{E}}

\def \sc {\text{sc}}

\def \v {\text{v}}
\def \f {\text{f}}
\def \b {\text{b}}
\def \sf {\text{sf}}

\def \l {\text{l}}

\begin{document}

\title{Nonlinear perturbations and weak shock waves in isentropic atmospheres}

\author{Tamar Faran}\email{tamar.faran@princeton.edu}\affiliation{Department of Astrophysical Sciences, Princeton University, Princeton, NJ 08544, USA}
\author{Christopher D.~Matzner}\affiliation{Department of Astronomy and Astrophysics, University of Toronto, 50 St. George Street, Toronto, ON M5S 3H4, Canada}
\author{Eliot Quataert}\affiliation{Department of Astrophysical Sciences, Princeton University, Princeton, NJ 08544, USA}

\begin{abstract}
Acoustic perturbations to stellar envelopes can lead to the formation of weak shock waves via nonlinear wave-steepening. Close to the stellar surface, the weak shock wave increases in strength and can potentially lead to the expulsion of part of the stellar envelope. While accurate analytic solutions to the fluid equations exist in the limits of low amplitude waves or strong shocks, connecting these phases generally requires simulations.
We address this problem using the fact that the plane parallel Euler equations, in the presence of a constant gravitational field, admit exact Riemann invariants when the flow is isentropic. We obtain exact solutions for acoustic perturbations and show that after they steepen into shock waves, Whitham's approximation can be used to solve for the shock's dynamics in the weak to moderately strong regimes, using a simple ordinary differential equation.
Numerical simulations show that our analytic shock approximation is accurate up to moderate ($\sim$ few--15) Mach numbers, where the accuracy increases with the adiabatic index.
\end{abstract}

\section{Introduction} \label{sec:intro}
The steepening of finite-amplitude sound waves into shocks is a foundational problem in nonlinear acoustics (\citealt{Riemann1860}, \citealt{salas2007curious}). It is also still of practical  interest in fields such as astrophysics (for stellar outbursts driven by core activity in evolved stars: \citealt{QuataertShiode12}, \citealt{McleySoker14_WaveDrivenExpansion}, \citealt{Fuller17_WaveHeatedRSGs}, \citealt{LinialFullerSari21}, \citealt{MatznerRo21apj}), solar physics (for the heating of the the Sun's corona: e.g., \citealt{Biermann46_ChromosphericSound,Ruderman06_SolarWaveHeating}), atmospheric physics (for the driving upper-atmosphere shocks from volcanic eruptions and rocket launches: \citealt{chou2018gigantic}), and laboratory fluid dynamics (for sonoluminesecence: \citealt{2001JFM...431..161L}).   

A common feature among these applications is that an amplifying effect, due to stratification or geometrical focusing, counters the damping effect of shock dissipation.   As a result even relatively weak disturbances have the potential to generate strong shocks.   However, a disturbance that does not carry enough energy may not evolve into a strong shock, either because the wave reflects rather than forming a shock at all, or due to energy dissipation during the weak phase of shock evolution.  

The limiting cases of this problem are informed by analytical solutions, especially for a wave well above the threshold for shock formation.  A weak disturbance amplifies, shears, and shocks as described by the conservation of Blokhintsev invariants \citep{blokhintsev1956acoustics} -- quantities related to the flow of energy and wave action along characteristics -- supplemented by the rules that govern weak shocks (such as the equal-area rule). A strong and accelerating shock, on the other hand, approaches the relevant similarity solution, such as \citet{sakurai60}'s solution for planar shocks in polytropic atmospheres. 

The transition between the weak and strong regimes is not well described by either limiting solution. Numerical solutions are generally required, therefore, to accurately determine the onset and properties of a strong shock solution from the initial conditions of the weak phase. Our goal in this work is to bridge the gap between the weak and strong regimes. To do that, we first obtain an exact solution for acoustic perturbations in equilibrium isentropic stellar atmospheres, by facilitating a modified version of the Riemann invariants in the presence of a constant gravitational field $g$. Next, we use the acoustic result to show that an approximate method, analogous to that of Whitham's \citep{Whitham58}, can be used to solve for the shock's dynamics.

The structure of this work is as follows. We begin by introducing the fluid equations, Riemann invariants and the hodograph transformation in \S~\ref{S:FluidEq}. In \S~\ref{S:AcousticSol} we solve for an acoustic perturbation of an initially static atmosphere, and apply Whitham's approximate method to obtain the dynamics of weak to moderately strong shock waves in \S\ref{S:Shock}.
We use numerical simulations to assess our analytic solutions' performance in \S\ref{S:Simulations} and conclude in \S\ref{S:Summary}.

\section{The fluid equations}\label{S:FluidEq}
Let us consider planar, {inviscid}, isentropic flow under the influence of a uniform acceleration field, described by the conservation of mass, momentum, and energy in one dimension. The fluid is assumed to have an adiabatic index $\gamma$, such that the adiabatic sound speed $c$ is related to the pressure, $p$, and the density, $\rho$, by $c^2 =\gamma p/\rho$. Under these assumptions, the continuity and momentum equations can be written as
\begin{subequations}\label{eq:FluidEq}
    \begin{equation}
        u_t+u\,u_x + \frac{2}{\gamma-1}c\,c_x = -g
    \end{equation}
    \begin{equation}
        c_t+u\,c_x + \frac{\gamma-1}{2}c\,u_x = 0\,,
    \end{equation}
\end{subequations}
where $u$  is the fluid velocity, $-g$ is a constant acceleration along the coordinate $x$ and subscripts denote partial derivative. 
It is straightforward to show that equations \eqref{eq:FluidEq} can be written in the following conservation form
 \begin{equation}\label{eq:Riemann_cons_general}
    \left[\partial_t+(u\pm c)\partial_x\right]J_\pm = 0\,.
\end{equation}
The quantities
\begin{equation}\label{eq:RiemannInv}
    J_\pm = u\pm \frac{2c}{\gamma-1}+gt\,
\end{equation}
are called the Riemann invariants and are conserved along curves of $dx = (u\pm c) dt$ that define $C_\pm$ characteristics.
When $u=0$ equations \eqref{eq:FluidEq} admit the relation of hydrostatic equilibrium:
\begin{equation}\label{eq:c0_eq}
    \frac{d c_0^2}{d x} = -g(\gamma-1) = -g/n\,,
\end{equation}
where subscript $_0$ refers to unperturbed quantities and we define
\begin{equation}
    n \equiv \frac{1}{\gamma-1}\,.
\end{equation}
In an isentropic medium $c^2\propto \rho^{1/n}$, and therefore $\rho_0 = k_\rho (-x)^n$ where the constant $k_\rho$ depends on entropy, and $x\leq0$.

The fluid equations can be reduced to a single partial differential equation (PDE) by performing a hodograph transformation.\footnote{A similar analysis with $g=0$ can be found in \S 105 of \citet{Landau}.} \citet{Pelinovskii1988} and \citet{Gundlach2009} have shown that by changing the independent variables to
\begin{equation}
    \lambda \equiv u+gt \,, ~~~~ \sigma \equiv \frac{2}{\gamma-1}c\,,
\end{equation}
equation set \eqref{eq:FluidEq} can be written as:
\begin{equation}\label{eq:PDE_Phi}
    \Phi_{\lambda\lambda}-\Phi_{\sigma\sigma}+\frac{\gamma-3}{\gamma-1}\frac{1}{\sigma}\Phi_\sigma=0\,,
\end{equation}
where $\Phi$ is related to $u$ and $x$ through its partial derivatives:
\begin{equation}\label{eq:v_sigma}
    u = \frac{1}{\sigma}\Phi_\sigma\,,
\end{equation}
and
\begin{equation}\label{eq:x}
    x = \frac{1}{2g}\left[(\gamma-1)\Phi_\lambda -u^2-\frac{\gamma-1}{2}\sigma^2\right]\,.
\end{equation}
Let us go back to Eq \eqref{eq:PDE_Phi} and make the transformation $\sigma = \sqrt{\Tilde{\sigma}}$. We then find
\begin{equation}\label{eq:PDE_SigmaTilde1}
\Phi_{\lambda\lambda}-2(1+2k)\Phi_{\Tilde{\sigma}}-4\Tilde{\sigma}\Phi_{\Tilde{\sigma}\Tilde{\sigma}}=0\,,
\end{equation}
where we define
\begin{equation}\label{eq:k_gamma}
   k = -\frac{\gamma-3}{2(\gamma-1)} = n-1/2 \,.
\end{equation}
Differentiating the above equation with respect to $\Tilde{\sigma}$, one finds that
\begin{equation}\label{eq:PDE_SigmaTilde}
(\Phi_{\Tilde{\sigma}})_{\lambda\lambda}-2\left[1+2(k+1)\right](\Phi_{\Tilde{\sigma}})_{\Tilde{\sigma}}-4\Tilde{\sigma}(\Phi_{\Tilde{\sigma}})_{\Tilde{\sigma}\Tilde{\sigma}}=0\,.
\end{equation}
Comparing equations \eqref{eq:PDE_SigmaTilde1} and \eqref{eq:PDE_SigmaTilde} shows that $\Phi_{k+1} = \partial_{\Tilde{\sigma}}\Phi_k$; i.e., if $\Phi_k$ is known for some value of $k$, solutions for higher integer values of $k$ can be derived from it via differentiation. The lowest order $k=0$ corresponds to $\gamma=3$.\footnote{Some integer values of $k$ correspond to more physically relevant values of the adiabatic index: e.g., $k=1$ corresponds to $\gamma=5/3$ and $k=2$ to $\gamma=7/5$.} 
Substituting $\gamma=3$ in Eq \eqref{eq:PDE_Phi} returns the wave equation $\Phi_{\lambda \lambda} = \Phi_{\sigma \sigma}$, whose general solution is
\begin{equation}\label{eq:Phi_k0}
    \Phi_0 = f_1(\lambda+\sigma)+f_2(\lambda-\sigma) = f_1(J_+)+f_2(J_-)\,,
\end{equation}
where $f_1$ and $f_2$ are arbitrary functions of $\lambda \pm \sigma = J_\pm$. The general solution for $k>0$ can be derived directly from $f_1(J_+)$ and $f_2(J_-)$, or alternatively represented as
\begin{multline}\label{eq:PhiGeneralSol}
    \Phi_k = \frac{1}{2^k}\left(\frac{1}{\sigma}\frac{\partial}{\partial \sigma}\right)^{k-1}\left(\frac{1}{\sigma}\frac{\partial \Phi_0}{\partial \sigma}\right) \\= \frac{1}{2^k}\left(\frac{1}{\sigma}\frac{\partial}{\partial \sigma}\right)^{k-1}\left[\frac{F_1(J_+)}{\sigma}+\frac{F_2(J_-)}{\sigma}\right]\,,
\end{multline}
where $F_1(J_+)$ and $F_2(J_-)$ are also arbitrary functions of $J_\pm$. Eq \eqref{eq:PhiGeneralSol} is identical to the result obtained by \citet{Landau} for the homogeneous fluid equations with $g=0$. It is also equivalent to the solution derived by \citet{Gundlach2009}.

The velocity, given by equations \eqref{eq:v_sigma} and \eqref{eq:PhiGeneralSol}, can be more conveniently written as a power series of $\sigma$:
\begin{equation}\label{eq:ExactVSol}
    u = \sum_{i = 0}^{k} \frac{F_1^{(i)} + (-1)^i F_2^{(i)}}{\sigma^{2k+1-i}} \times P(i)\,,
\end{equation}
where
\begin{equation}
    P(i) = \begin{cases}
         \prod_{j=0}^{k-1-i}\left[\frac{j}{2}-\frac{k(k+1)}{2(j+1)}\right]&  ~,~   i<k \\
         1  & ~,~ i=k
        \end{cases}
\end{equation}
and parenthesized superscripts denote the number of derivatives with respect to the argument $J_+$ or $J_-$.
To leading order in $\sigma^{-1}$ (as  $\sigma \rightarrow\infty$), the velocity is
\begin{equation}\label{eq:uSol_leading_order}
    u \simeq\frac{F_1^{(k)}}{\sigma ^{k+1}} + (-1)^k\frac{F_2^{(k)}}{\sigma ^{k+1}}\,,
\end{equation}
i.e., $u \propto \sigma^{-k-1} = \sigma^{-n-1/2}$  along a characteristic.  
This limit agrees with linear theory \cite[e.g.,][]{Lighthill} in which $\rho u^2 \sigma A$ is conserved along one set of characteristics, where $A$ is the cross section.

In the next section, we use the above solution with appropriate boundary conditions to obtain the dynamics of an acoustic perturbation in a medium that is initially in a state of hydrostatic equilibrium.

\section{Acoustic perturbation}\label{S:AcousticSol}
In this section we solve for an acoustic perturbation to an isentropic atmosphere that is initially in hydrostatic equilibrium. The initial width of the perturbation is assumed much smaller than the local scale height, namely, $\Delta x \ll x_\i$.
The static fluid at $x>x_\i$ is perturbed upon the arrival of a limiting $C_+$ characteristic that originates from the front end of the perturbed region and carries a constant value of the Riemann invariant, designated by $J_{+_ l} = (\lambda +\sigma)_ l = a$. Since the limiting characteristic defines the boundary between the perturbed and the unperturbed regions, the boundary conditions along its path are $u = 0$ and $\sigma = \sigma_0$,  enforcing $\Phi_\lambda = \Phi_\sigma = 0$ along $J_+=a$, according to equations \eqref{eq:v_sigma} and \eqref{eq:x}. We find that for an arbitrary initial profile (determined by the choice of $F_1$), the following form of $F_2$ satisfies the desired boundary conditions:
\begin{equation}\label{eq:F2Sol}
    F_2 = - \sum_{\alpha=0}^{k}(J_--a)^{\alpha} F_1^{(\alpha)}(a) \, P(\alpha)\,S(\alpha)^{-1}\,,
\end{equation}
where
\begin{equation}
    S(\alpha) = \sum_{i=0}^{\alpha} \frac{\alpha!}{(\alpha-i)!} 2^{\alpha-i}(-1)^{\alpha} P(i)\,.
\end{equation}
We note that this solution does not take into account an initial stage in which $C_+$ and $C_-$ characteristics from within the wave profile intersect. However, since the initial width of the perturbation is assumed to be small, that stage is very brief and is thus of little interest. Moreover, the solution is valid only until the limiting characteristic arrives at $x=0$ and reflects backwards into the perturbed medium. We do not treat this phase in this work.
\newline

\begin{figure}
    \centering
    \includegraphics[width = 0.44\textwidth]{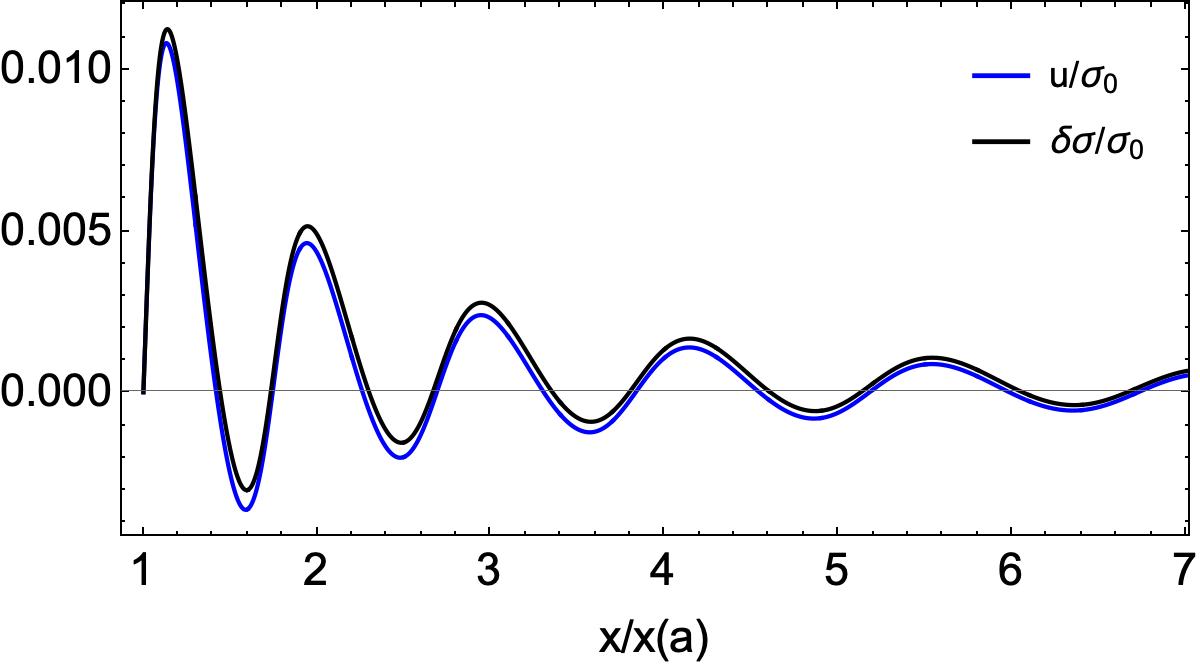}
    \includegraphics[width = 0.47\textwidth]{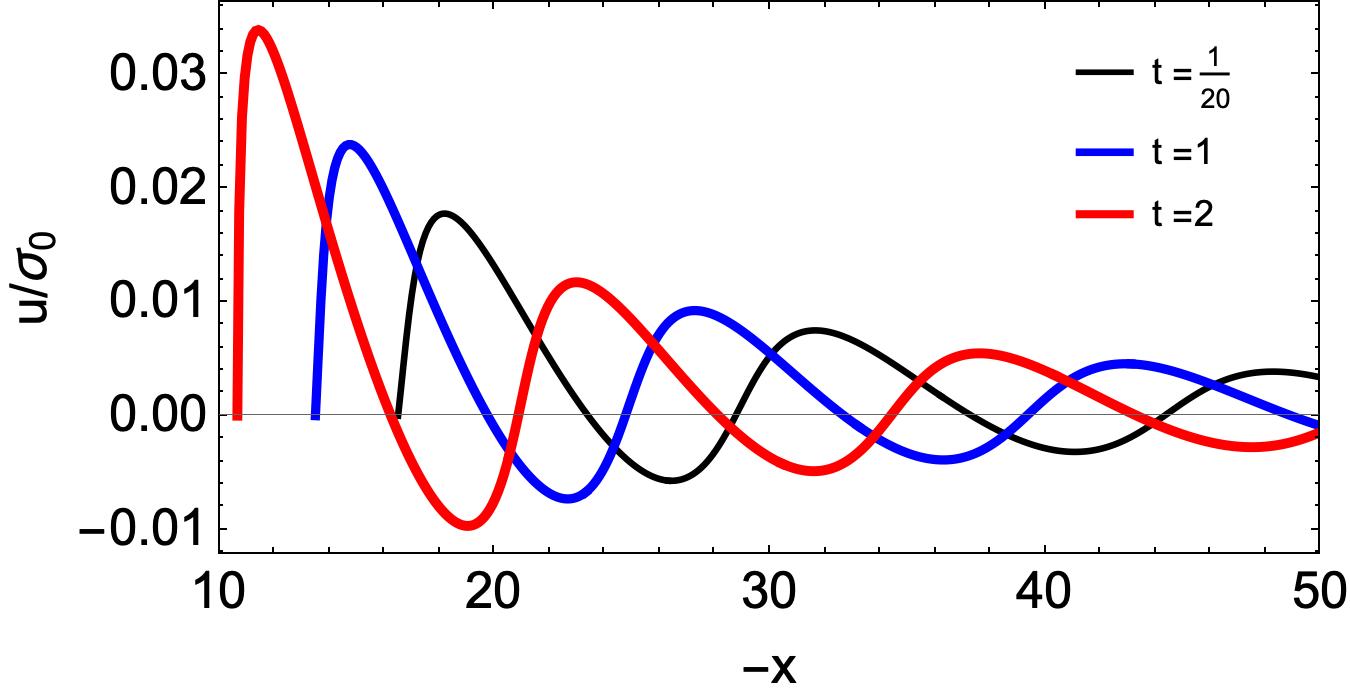}
    \caption{ Visualization of the acoustic solution given by Eqs \eqref{eq:ExactVSol} and \eqref{eq:F2Sol} for $\gamma = 5/3$ $(k=1)$, with a choice of $F_1 = A \cdot Sin[\frac{\lambda + \sigma}{B}]$, using numerical values of $g = 1$, $a=10$, $A = -5$ and $B=1/2$. Top: the relative strength of the perturbations to $u$ and $\sigma$ at time $t = 1/20$. The profile terminates at the location of the limiting $C_+$ characteristic, $x(a)$. Bottom: the relative velocity perturbation at different times. The acoustic perturbation steepens into a shock as it approaches the surface.}
    \label{fig:acoustic}
\end{figure}

\autoref{fig:acoustic} depicts the acoustic solution for the case $\gamma = 5/3$ $(k=1)$ using equations \eqref{eq:ExactVSol} and \eqref{eq:F2Sol} for one representative choice of $F_1(J_+)$. Contrary to linear theory, it is evident that in general $u \neq \sigma-\sigma_0$.

\subsection{The special case of $\gamma=3$}
The dynamics of Eq \eqref{eq:Riemann_cons_general} are unique when $\gamma=3$ $(n =1/2)$, as already implied by the results of the previous section. Since $\sigma = c$ in this special case, Riemann invariants carry a constant propagation velocity in a freely-falling reference frame, namely
$J_\pm = u'\pm c = u+gt\pm c$, where $u' = u+gt$ is the fluid velocity in the free-fall frame. This quality suggests that characteristics of different families do not interact, as expected from the solutions of the wave equation (Eq \ref{eq:Phi_k0}).

The boundary conditions in the case of $k=0$ require that $F_2(J_-)=0$, since the envelope is unperturbed prior to the arrival of the limiting $C_+$ characteristic.
The velocity therefore satisfies
\begin{equation}\label{eq:VExactGamma3}
    u = \frac{F_1(J_+)}{\sigma}\,,
\end{equation}
so that the product $u \cdot \sigma$ is constant along $C_+$ characteristics. Substituting Eq \eqref{eq:VExactGamma3} in Eq \eqref{eq:x}, one finds that the difference in $\sigma$ with respect to the unperturbed state is
\begin{equation} \label{eq:deltasigmaGamma3}
    \delta \sigma \equiv \sigma-\sigma_0 = u\,.
\end{equation}
The velocity along a $C_+$ characteristics can now be written as a function of $c_0 = \sigma_0$:
\begin{equation} \label{eq:v_SW_gamma3}
 u = \pm \left[ u_\i c_\i + (c_0/2)^2 \right]^{1/2} - c_0/2\,,
\end{equation}
where the positive (negative) branch in each case represents positive (negative) velocities, and the values of $u_\i$ and $c_\i$ correspond to the initial values on the $C_+$ characteristic along which the solution holds. Eq \eqref{eq:v_SW_gamma3} is equivalent to Eq \eqref{eq:VExactGamma3} where $F_1(J_+) = u_\i c_\i$.

\section{Approximate shock dynamics}\label{S:Shock}
The wave steepens as it propagates through the stellar atmosphere and a discontinuity may form if $u_x \rightarrow \infty$ at some point in its profile before it reaches $x=0$.
\citet{Gundlach2009} showed that $u_x$ and $\sigma_x$ can be written as
\begin{equation}\label{eq:vx_sigmax}
    u_x = \frac{1}{g \Delta}u_\sigma\,,~ \sigma_x = \frac{1}{g \Delta}(1-u_\lambda)
\end{equation}
where $\Delta$ is the Jacobian of Eq \eqref{eq:PDE_Phi}:
\begin{equation}\label{eq:Jacobian}
    \Delta = -\frac{\sigma}{2 n g^2}\left[(1-u_\lambda)^2-u_\sigma^2\right]\,.
\end{equation}
A discontinuity forms when $\Delta = 0$, which depends on the initial properties of the waveform. The discontinuity then proceeds as a shock wave, and since energy is dissipated across the shock front, the flow is no longer isentropic. 
Nevertheless, expansion of the shock jump conditions in the limit of small perturbations shows that the leading term in the entropy jump $\Delta S$ is only of third order in the small parameter $\delta \equiv\Delta p/p_\u$:
\begin{equation}\label{eq:DeltaS}
    \Delta S \simeq c_\v\frac{\gamma^2-1}{12 \gamma^2}\delta^3 = \frac{\gamma+1}{12 \gamma^2}\mathcal{R}^{-1}\delta^3\,,
\end{equation}
where $\Delta p = p_\d-p_\u$ is the jump in pressure across the shock, $S = c_\v \log P/\rho^\gamma$ is the entropy and subscripts $_\d$ and $_\u$ denote immediate downstream and upstream values, respectively. We also used the expression for the specific heat at constant volume for polytropic gases, $c_\v = \frac{\mathcal{R}}{\gamma -1}$, where $\mathcal{R}$ is the gas constant.
In the same way, it can be shown that the lowest order term in the relative change in $J_-$ is also of third order in the shock strength. Since we are interested in the transition between weak and strong shock waves, it is reasonable to assume that the shock is effectively isentropic. In this limit, $\Delta \sigma \simeq \Delta u$ across the shock front \cite[e.g.,][]{whitham1974linear}.

The conservation of Riemann invariants along $C_+$ characteristics can be employed to solve for the dynamics of the shock under some simplifying assumptions. \citet{Whitham58} has demonstrated that approximate solutions can be obtained by applying the shock jump conditions along the differential characteristic relation. The Riemann invariant is assumed uniform in the post-shock flow, rather than just constant along a specific $C_+$ characteristic. This approximation implicitly assumes that perturbations to the shock front are localized, and are attributed to local changes in the density or the cross section of the upstream fluid. Interaction with post shock flow, such as reflected signals, plays a negligible role in dictating the shock's dynamics.
The assumption is supported by studies such as that of \citet{Chisnell1957}, who showed that reflected signals that are re-reflected towards the shock have an overall cancelling effect, so that their collective contribution (after one re-reflection) is negligible compared to their individual effect. When all re-reflected signals are neglected, the shock is described by Whitham's approximate equation.
Although only partial justification has been provided for the use of this approximation, the method has been proven successful when applied to various problems of shock propagation.
A notable example, related to the problem considered in this work, is the remarkable agreement with \citet{sakurai60}'s exact self-similar solution for a strong shock wave that accelerates near the surface of a stellar atmosphere.

As explained above, we are interested in shocks of weak to moderate strength, for which energy dissipation can be neglected, and as a consequence, so can any reflected signals generated at the shock. In this limit, we can expect that assuming localized perturbations is especially adequate. We will now use the acoustic solution found in the previous section to provide further justification for using the approximation by showing that $J_+$ is approximately constant in space in the post shock fluid.

Along the shock's trajectory, the following equality holds:
\begin{equation}\label{eq:RiemannInvShock}
    u_\d+gt+\sigma_\d = (u+gt +\sigma)_\i \,,
\end{equation}
where subscript $_\i$ labels the initial properties of the $C_+$ characteristic that overtakes the shock at time $t$. The reference time $t_\i$ is constant for all characteristics. At sufficiently late times, characteristics arriving at the shock originate at the base of the initial wave profile, where $u_\i \ll \sigma_{0,\i}$ and therefore $\sigma_\i \simeq \sigma_{0,\i}$. 
One can expect that as the shock approaches $x \rightarrow 0$ and accelerates, characteristics from a narrowing region within the initial profile are able to overtake it, such that $J_{+,\i} \simeq \sigma_{0,\i} +g t_\i \simeq const$. It is easy to justify this assumption in the case of $\gamma=3$, since the solution depends only on $F_1(J_+)$. Using equations \eqref{eq:VExactGamma3}, \eqref{eq:vx_sigmax} and \eqref{eq:Jacobian} we find the spatial change in $J_+$ at constant time:
\begin{equation}
\begin{split}
    \partial_x(u+gt+\sigma) &= -\frac{2 n g}{\sigma(1-u_\lambda-u_\sigma)} \\ &= -\frac{g}{\sigma -2 F_1'(J_+)+F_1(J_+)/\sigma}\,.
    \end{split}
\end{equation}
Using this result and the fact that $\sigma \simeq \sigma_0 + u$ across the shock, the relative change in $J_+$ across a typical scale $x$ is
\begin{equation}\label{eq:delJp_Jp}
    \frac{\Delta J_+}{J_+} =\frac{\partial_x J_+ \cdot x}{J_+} = \frac{1}{2}\frac{\sigma_0}{\sigma_{0,\i}}\frac{1}{1+2u/\sigma_0-2 F_1'(J_+)/\sigma_0}\,.
\end{equation}
In order to have $\Delta J_+/J_+ \rightarrow 0$ as $\sigma_0 \rightarrow 0$, the last term in the denominator of Eq \eqref{eq:delJp_Jp} must be negligible, i.e., $|F_1'(J_+)|/\sigma_0 \ll 1$. If we rewrite $F_1(J_+) = u_\i(J_+-u_\i-g t_\i)$ then $F_1'(J_+) = u_\i+u_\i'(J_+-2u_\i-g t_\i) = u_\i + u_\i' \sigma_{0\i}$, where primes denote derivatives with respect to $J_+$. In the limit where $u_\i \ll \sigma_{0\i}$ and therefore $J_+ \simeq \sigma_{0\i}+g t_\i$, the requirement can be stated as $|d u_\i/d \sigma_{0\i}| \ll \sigma_0/\sigma_{0\i} = \sqrt{x/x_\i}$, recalling that $g t_\i$ is a constant. Given an initial velocity profile that decreases fast enough as to satisfy that condition, $J_+$ is effectively uniform in space as $x\rightarrow0$, and it is justified to assume the constancy of $J_+$ along the shock as it propagates in that region.

Although we have not justified our assumption for any arbitrary value of $\gamma$, we expect it to hold in general as long as perturbations carried along $C_-$ characteristics are small [which depends on $F_2(J_-)$], which is a reasonable assumption for sufficiently weak shock waves.

Setting $\Delta \sigma = u$ across the shock front, Eq \eqref{eq:RiemannInvShock} can be written as $2u_\d + g(t-t_\i) + \sigma_0 = \sigma_{0,\i}$.
Switching to dimensionless variables
\begin{equation}
    \tau \equiv (t-t_\i)/t_\sc \,,~~~m_\d \equiv u_\d/c_0\,,
\end{equation}
where $t_\sc = \sigma_{0,\i}/g$ is the initial sound crossing time, we have
\begin{equation}\label{eq:ShockChar}
    \frac{\sigma_0}{\sigma_{0,\i}}\left(m_\d/n+1\right)+\tau = 1\,.
\end{equation}
Next, we take the advective derivative of Eq \eqref{eq:ShockChar} along the shock trajectory and obtain the following ordinary differential equation (ODE):
\begin{equation}\label{eq:shock_ODE}
    \frac{d m_\d}{d\tau} = \frac{\left[M_\sh(m_\d/n+1)-1\right](m_\d+n)}{1-\tau}\,,
\end{equation}
where we used $d(\sigma_0/\sigma_{0,\i})/d\tau = -M_\sh$. Eq \eqref{eq:shock_ODE} can alternatively be represented as an integral equation
\begin{equation}\label{eq:shock_integral_eq}
    \int_{m_{\d0}}^{m_\d}\frac{dm}{(m+n)\left[1-(1+m/n)M_\sh\right]} = \ln\left[\frac{1-\tau}{1-\tau_0}\right]\,,
\end{equation}
where $\tau_0$ is some initial time and $m_{\d0} = m_\d(\tau_0)$.
It still remains to find $M_\sh(m_\d)$. 
It is valid to use either the exact relation 
$M_\sh-1/M_\sh = (\gamma+1)m_d/2$
(assuming $u_u=0$), or an approximate relation consistent with the assumption of constant entropy. 
We take the latter approach, as it is appropriate to sufficiently weak shock waves and as it makes $M_\sh(m_\d)$ analytical.
Conservation of mass flux across the shock front implies that its velocity is $U_\sh = \frac{\rho_\d u_\d-\rho_\u u_\u}{\rho_\d-\rho_\u}$. Using the fact that $\Delta \sigma \simeq \Delta u$ at the shock and assuming $u_\u = 0$, we find
\begin{equation}\label{eq:jump_conditions}
    M_\sh = \frac{m_\d}{1-\left(1+\frac{m_\d}{2n}\right)^{-2n}}\,,
\end{equation}
which can now be substituted into Eq \eqref{eq:shock_ODE}.

\begin{figure}
    \centering
    \includegraphics[width = 0.44\textwidth]{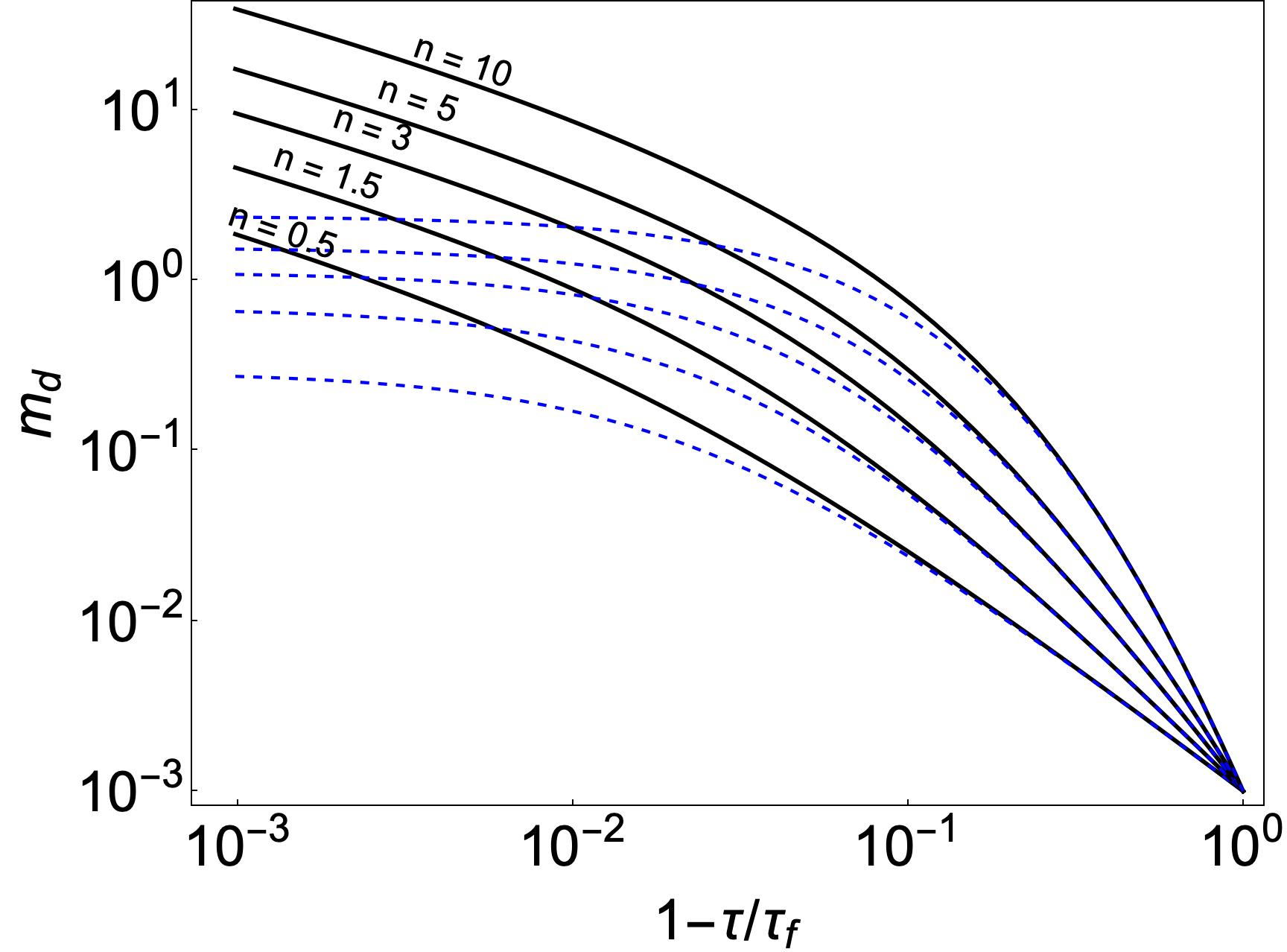}
    \caption{The Mach number of the fluid in the immediate downstream of the shock for various values of $n$, as obtained from solving Eq \eqref{eq:shock_ODE} with the initial condition $m_{\d0} = 10^{-3}$. The evolution is shown as a function of $1-\tau/\tau_\f$, where $\tau_\f$ is the time the shock reaches $x=0$. The acceleration of the shock wave increases with $n$ and $m_\d$ diverges at shorter times. The Dashed lines are linearized solutions of Eq \eqref{eq:shock_ODE}.}\label{fig:msh_ODE}
\end{figure}

Eq \eqref{eq:shock_ODE} can be numerically integrated up to $\tau = \tau_\f$, defined as the time the shock reaches $x = 0$. This point is a movable singularity of the equation, whose location depends on the initial conditions and on the value of $n$. Given a solution for $M_\sh(\tau)$, the shock's position can then be found using the following relation
\begin{equation}
    \frac{x}{x_0} = \left[1-\int_{\tau_0}^\tau M_\sh(\tau')d\tau'\right]^2\,.
\end{equation}
An interesting property of the solution is the existence of two characteristic times, corresponding to $\tau = 1$ and $\tau = \tau_\f$. The two are comparable when $n$ is of order unity, since the shock spends most of its time crossing the initial scale-height.
In Figure \ref{fig:msh_ODE} we show the solution to Eq \eqref{eq:shock_ODE} for various values of $n$, using the same initial condition, $m_\d = 10^{-3}$ at $\tau_0 = 10^{-5}$.

\subsection{Analytic shock solution for $\gamma=3$}
When $\gamma=3$, the criterion for shock formation, $\Delta =0$, involves only derivatives of $F_1$:
\begin{equation}
    \left(1-\frac{2 F_1'}{\sigma}+\frac{F_1}{\sigma^2}\right)\left(1-\frac{F_1}{\sigma^2}\right) = 0\,,
\end{equation}
which is solved by either $u = \sigma$ or $u+\sigma =2 F_1'$, corresponding to $(J_-)_x \rightarrow\infty$ and $(J_+)_x\rightarrow \infty$, respectively.
The first solution implies $\sigma_0=0$ and is therefore trivial. The second condition together with Eq \eqref{eq:v_SW_gamma3} show that a discontinuity never forms if
\begin{equation}
    F_1' = \frac{u+\sigma}{2} = \left[u_\i \sigma_\i + (\sigma_0/2)^2\right]^{1/2} \,.
\end{equation}
Therefore, if everywhere on the pulse's profile $F_1'<(u_\i \sigma_\i)^{1/2}$, a discontinuity cannot form; instead, the wave will reflect without creating a shock.
If a shock does form before the pulse arrives at the stellar surface, its dynamics are described by Eq \eqref{eq:shock_ODE}, which can be integrated analytically:
\begin{equation}
    \left(\frac{m_d}{m_{\d0}}\right)^{-2/3}\left(\frac{3+2m_d}{3+2m_{\d0}}\right)^{-1/3}\frac{1+2m_d}{1+2m_{\d0}} = \frac{1-\tau}{1-\tau_0}\,.
\end{equation}
Rearranging, we find an explicit expression for $m_\d(\tau)$:
\begin{equation}\label{eq:md_gamma3}
    m_\d(\tau) = \frac{1}{2}\left[\zeta F(\zeta)^{-1}+F(\zeta)(\zeta-1)^{-1}-1\right]\,,
\end{equation}
where
\begin{equation*}
    \zeta \equiv  \frac{(1+2m_{\d0})^3 } 
    {4m_{\d0}^2 (3+2m_{\d0})}\left(\frac{1-\tau}{1-\tau_0}\right)^3\,
\end{equation*}
and
\begin{equation*}
    F(\zeta) = \left[-\zeta(\zeta-1)^2+\zeta(1-\zeta)^{3/2}\right]^{1/3}\,.
\end{equation*}
The time it takes the shock to travel to $x=0~ (m_\d \rightarrow \infty)$, can be inferred from the singularities of Eq \eqref{eq:md_gamma3}. The singular point at $\zeta = 0$ is trivial, as it is equivalent to $\tau = 1$. This point lies outside of the solution's domain since the shock always travels faster than the initial sound speed, and therefore the solution terminates at $\tau<1$. The relevant singular point is $\zeta=1$, corresponding to an arrival time at the surface of
\begin{equation}
    \tau_\f = 1- \frac{(1-\tau_0)}{1+2m_{\d0}}\left[4m_{\d0}^2(3+2m_{\d0})\right]^{1/3}\,.
\end{equation}
As expected, $\tau_\f$ becomes shorter as the initial Mach number of the shock increases. 

\section{Numerical Simulations}\label{S:Simulations}
We test our analytical results of \S \ref{S:Shock} against numerical simulations using a code provided by  Elad Steinberg (private communication). The simulation uses a second order Lagrangian Godunov scheme and an exact Riemann solver with a minmod slope limiter {to solve hydrodynamics in planar symmetry}.
We initialize the grid with $10^5$ cells, divided logarithmically {in depth} over the range between a lower boundary $z_{\rm in} = 0.5$ and an upper boundary $z_{\rm ext}=1-10^{-4}$, where the surface is set to $R=1$ and $z \equiv R-|x|$ is the distance from the surface. The slight offset is required for stability, and we apply a reflecting boundary at $z_{\rm ext}$ to balance the external pressure; this does not affect the dynamics of interest. We consider a uniform gravitational acceleration, where the functional forms of the initial density and pressure profiles satisfy hydrostatic equilibrium for a light envelope, with $\rho_0 = (2z/R)^n$ and $P_0 =\frac{g R}{n+1}(2z/R)^{n+1}$, such that the density in the internal boundary is chosen to be $1$ and the pressure at the surface is $0$.  We find that numerical deviations from hydrostatic equilibrium decrease with the grid's resolution, as expected. We run the simulation for three different density profiles, corresponding to $\gamma = 4/3 ~(n=3)$, $\gamma = 3/2~ (n=2)$ and $\gamma = 5/3~ (n=3/2)$. For each case, we superpose a Gaussian perturbation in velocity centered at $z/R \simeq 0.45$ with a Mach number amplitude $m_\p = 0.01$ and typical widths of $\Delta z \simeq 10^{-3}$ for $n=3$, and $\Delta z \simeq 0.002$ for $n=2$ and $n=3/2$. We perturb $\rho_0$ and $P_0$ accordingly, to maintain $v = \delta \sigma$ and minimize backwards travelling signals.
We verify that the grid's resolution is sufficiently high, so that the effect on the shock's dynamics is negligible.

\begin{figure*}
    \centering
    \includegraphics[width = 0.7\textwidth]{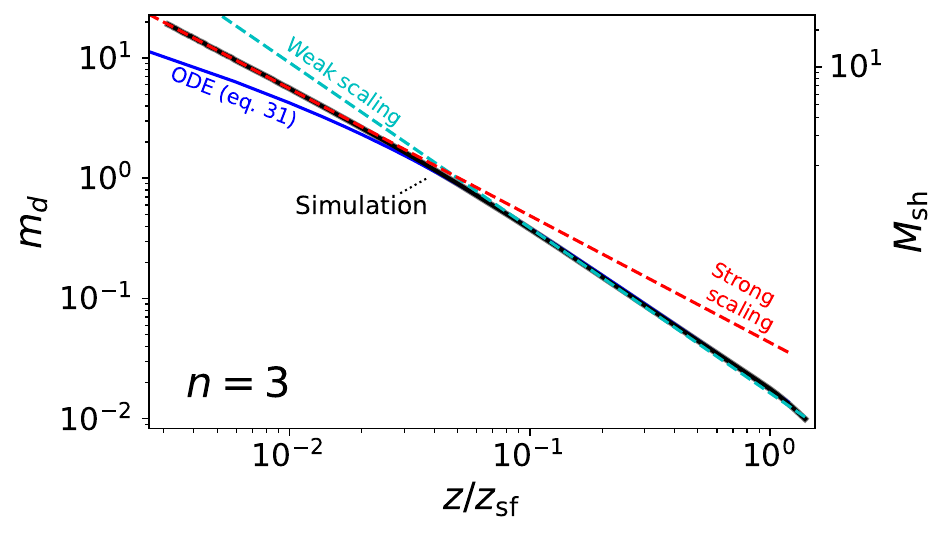}
    \includegraphics[width = 0.7\textwidth]{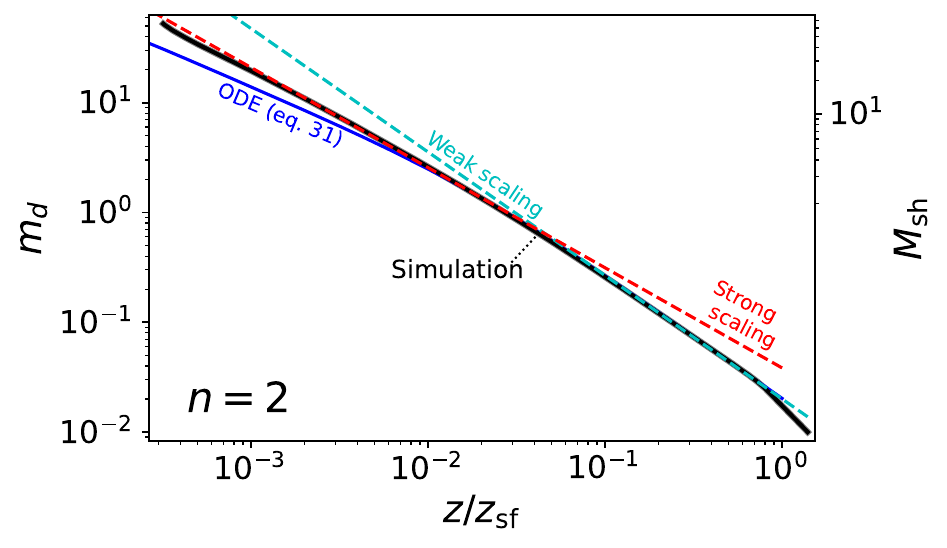}
    \includegraphics[width = 0.71\textwidth]{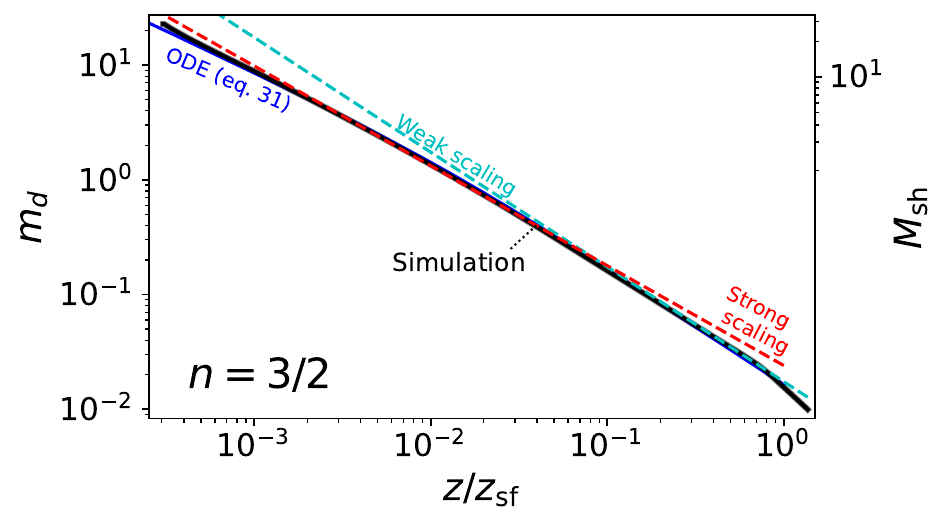}
    \caption{Numerical simulations for the dynamics of an acoustic pulse propagating down a background medium initially in hydrostatic equilibrium, for 3 different values of $\gamma$; top: $\gamma = 4/3 (n=3)$, middle: $\gamma = 3/2 (n=2)$, bottom: $\gamma = 5/3 (n=3/2)$.  Here $z = -x$, and $z_{\sf}$ is the location of shock formation. The $y$ axis on the left corresponds to the relative Mach number ($v/c_0$) of the fluid at the peak of a pulse, which is also the Mach number of the fluid in the downstream of the shock after it has matured and became an $N$-wave. On the right, we show the shock Mach number, and the $x$ axis is the distance from the medium's surface, normalized to the location of shock formation. We compare against our analytical model of Eq \eqref{eq:shock_ODE} (blue solid line). The limiting scalings of strong \cite{sakurai60} and weak shock dynamics are plotted as dashed lines. The agreement of Eq \eqref{eq:shock_ODE} with the numerical solution extends to higher Mach numbers as $n$ decreases, owing to the dependence of the entropy jump across the shock on $\gamma$. We note that the strong shock limit is not reached within the domain of our simulation in the case of $n=3/2$ and $n=2$.}\label{fig:Msh_sim}
\end{figure*}

In Figure \ref{fig:Msh_sim} we plot the peak of $v/c_0$ as a function of the distance from the surface, normalized to the location of shock formation. As the shock matures and evolves into an N-wave, the peak of the profile coincides with $m_\d$. 

For $n=3$, we see that once the shock matures, its dynamics agree well with the solution of Eq \eqref{eq:shock_ODE}, up to $m_\d \sim 2.5$, beyond which \citet{sakurai60}'s self-similar solution ($M_\sh \propto (-x)^{-\beta\cdot n-1/2}$, where the value of $\beta$ is taken from \citealt{Ro2013}), becomes a better description of the dynamics. We also plot the linearized weak shock scaling, $m_\d \propto (-x)^{-\frac{1}{4}(n+5/2)}$, which departs from the numerical solution at $m_\d \sim 1$.

The agreement between our isentropic shock model and the numerical results extends to higher Mach numbers as $n$ decreases together with the entropy jump at the shock, as can be seen for $n=2$ and $n = 3/2$ in the middle and bottom panels of Figure \ref{fig:Msh_sim}. 
Excellent agreement is obtained for $n=3/2$, where our isentropic shock model traces the numerical results up to $m_\d \sim 10$ ($M_\sh \sim 13$). In this case, Sakurai's strong shock solution is not reached within our simulation domain.

\begin{figure}
    \centering
    \includegraphics[width=0.92\columnwidth]{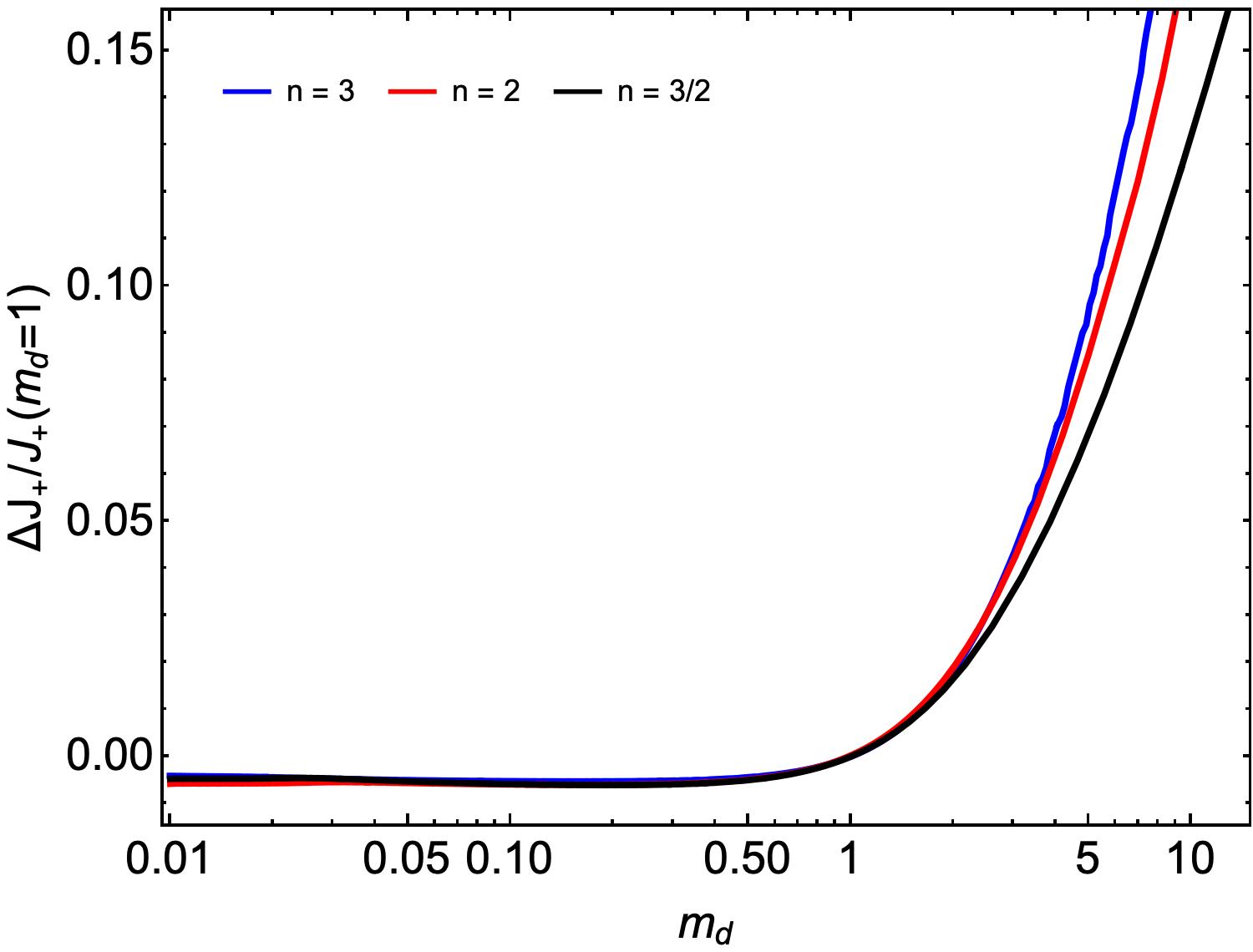}
    \caption{The relative change in the value of $J_+$ at the shock's position with respect to its value when $m_d = 1$, as computed from the numerical results of \S \ref{S:Simulations}. $J_+$ is effectively constant along the shock in the weak regime, but starts changing significantly as the shock becomes strong. The accuracy of Whitham's approximation is better for smaller values of $n$, i.e., higher values of $\gamma$.}
    \label{fig:JpShock}
\end{figure}

The agreement of the numerical results with the underlying assumption of our model is seen in Figure \ref{fig:JpShock}; the value of $J_+$ along the shock's path is effectively constant before entering the strong regime. As expected, the accuracy of the assumption increases with $\gamma$ (decreases with $n$).

\section{Summary and Discussion}\label{S:Summary}
We use the existence of Riemann invariants in isentropic, gravitationally-stratified atmospheres to obtain exact nonlinear solutions of acoustic perturbations assuming a polytropic equation of state. In the small amplitude limit, our solutions agree with the limiting relations of linear theory in which wave action is conserved \citep{blokhintsev1956acoustics,Lighthill}.

The pulse profile deforms as it propagates through the medium and can develop a discontinuity before reaching its surface, resulting in the formation of a shock wave. Beyond this point, the assumption of adiabaticity is no longer exact due to dissipation of energy across the shock front. Nevertheless, the fact that the jump in entropy and in $J_-$ are only of third order in the shock's strength allows us to neglect changes to the medium's entropy and use the conservation of $J_\pm$ to study the dynamics of the shock up to moderate Mach numbers.

Using our acoustic solution, we show that in the limit $x\rightarrow 0$ the value of $J_+$ is effectively uniform in space, which allows us to apply the conservation of $J_+$ along the shock's path in order to infer its dynamics. This method is analogous to \cite{Whitham58}'s well known approximation.
Conservation of mass flux across the shock front is used to derive the shock jump conditions, accurate to second order in $\Delta p/p_\u$, from which we obtain a simple ODE for the fluid's Mach number $m_\d$ as a function of the dimensionless time, $\tau$. In these variables, the ODE is scale free and depends only on the value of $n$. The solution terminates at a movable singularity of the equation, which depends on the initial conditions and corresponds to a time $\tau_\f<1$ at which the shock reaches the surface of the medium and attains an infinite Mach number. 

Some studies, e.g., \citet{Lighthill} and \citet{MatznerRo21apj} obtain an approximate shock solution using a modified equal area rule. The most notable difference between our solution and previous works is the accuracy of the jump conditions applied across the shock. While both methods use $\Delta \sigma = \Delta u$, which is accurate to second order in $\Delta p/p_\u$, the equal area rule also uses $U_\sh \simeq \frac{1}{2}\left[(u_\u+c_\u)+(u_\d+c_\d)\right]$, which is only accurate to linear order in $\Delta p/p_\u$.

The accuracy of our isentropic shock solution increases with the adiabatic index: since the jump in entropy is a decreasing function of $\gamma$, the change to the medium's entropy is smaller for shallower density profiles.
To test these predictions, we run numerical simulations for three different values of $n$ of an initially weak acoustic pulse, propagating down an unperturbed atmosphere. The numerical results shown in Figure \ref{fig:JpShock} confirm our expectation; the agreement between the solution of Eq \eqref{eq:shock_ODE} and the simulation extends to higher Mach numbers as $n$ decreases. We find that for $n=3$ ($\gamma = 4/3$), significant deviation between the analytic solution and the numerical results occurs already at $M_\sh \sim 4$, while for $n=3/2$ ($\gamma = 5/3$) the numerical and isentropic solutions agree at least up to $M_\sh \sim 20$. Once the entropy jump becomes important, our solution deviates in behavior from Sakurai's self-similar solution for strong shocks.

The special case of $\gamma=3$ ($n=1/2$) is studied in detail. This case is unique as $C_\pm$ characteristics retain a constant propagation velocity in the freely-falling frame, and accelerate at a constant rate, $-g$, in the lab frame, tracing parabolic, ballistic trajectories. We present an analytic solution for the entire evolution of the pulse, in both the acoustic and shock wave regimes.

The results of this paper can be directly applied to sub-energetic supernova explosions, in which the prompt shock wave is not strong enough to traverse the entire stellar envelope and unbind it. It has been shown by \cite{Lovegrove2013} and  \cite{Fenandez18} that some mass can still be lost due to the hydrodynamic response of the outer envelope to an instantaneous loss of $\sim 0.2-0.5 M_\odot$ to neutrinos during the protoneutron star phase of core collapse. The readjustment of the envelope to the modified gravitational potential creates an acoustic perturbation, which may steepens into a shock wave and unbind part of the external envelope. A reliable prediction of the amount of mass lost in the process and the accompanying electromagnetic signal requires a careful analysis of the evolution from an acoustic perturbation to a strong shock.

\section*{Acknowledgements}
We thank Elad Steinberg for sharing his Lagrangian 1D hydrodynamical code and Andrea Antoni for assisting with the numerical setup. CDM's research was supported by an NSERC Discovery Grant.    This research benefited from interactions at workshops funded by the Gordon and Betty Moore Foundation through Grant GBMF5076. TF is grateful to the CCPP department at NYU for hosting her as a visiting scholar.

\bibliography{WeakShock}{}
\bibliographystyle{aasjournal}

\end{document}